\documentclass[conference]{IEEEtran}

\usepackage{amsmath,amssymb,cite}
\usepackage{graphicx} 
\usepackage{epstopdf}

\newcommand{\id}{\mathrm{id}}

\newcommand{\benum}{\begin{enumerate}}
\newcommand{\eenum}{\end{enumerate}}

\newcommand{\ket}[1]{\left| #1 \right\rangle}

\newcommand{\proj}[1]{\left| #1 \right\rangle \left\langle #1 \right|}

\newcommand{\phip}{\phi^{+}}

\begin{document}

\title{Addressing  a device in a quantum network: A quantum approach including routing}

\author{
\IEEEauthorblockN{Alexander Pirker\IEEEauthorrefmark{1}\IEEEauthorrefmark{2}}
\vspace{0.05in}
\IEEEauthorblockA{\IEEEauthorrefmark{1}Universität Innsbruck, Institut für Theoretische Physik, Technikerstraße 21a, 6020 Innsbruck, Austria}
\vspace{0.05in}
\IEEEauthorrefmark{2}Quantum Network Design GmbH, Technikerstraße 21a, 6020 Innsbruck, Austria}

\date{\today}

\IEEEtitleabstractindextext{%
    \begin{abstract}
        In this work we propose an addressing scheme for quantum networks which relies on quantum states held by devices. Quantum network devices use their address state together with a request state that encodes the tasks to be executed. Our approach not only removes the necessity to classically communicate addresses, but also the need to communicate the operations a device must apply. It turns out that utilizing entanglement to encode addresses of devices in a quantum network leads to interesting applications such as overlaying different network states. We present a distributed quantum routing protocol using entanglement that coherently selects a route in a network of Bell-states for controlled-teleportation and lastly we prove that addressing using quantum states is equivalent to performing tasks in superposition in a quantum network. 
    \end{abstract}
}

\maketitle
\IEEEdisplaynontitleabstractindextext

\section{Introduction}

Quantum information and its applications caused a hype in many industry sectors. Quantum computing holds the promise to solve many tasks intractable for classical computers such as integer factorization \cite{ShorFactor}. It is also believed that applications like distributed quantum computing \cite{Caleffi18a,Cacciapuoti18} or quantum key distribution \cite{ShorQKD} will revolutionize current information technologies. At the very heart of many of those applications are quantum networks and the quantum internet \cite{KimbleQInet}. 

A quantum network \cite{Pirker18, VanMeter2014book} corresponds to a network of quantum devices. Those may offer several services. For instance, a quantum network could offer a service to send qubits from one device to another but also connect quantum computers via entanglement \cite{Meter2011,Miguel-Ramiro2021}. A network could also serve for key distribution. Research on quantum networks is highly active in many areas \cite{Hayashi2007,Meter2013a,Pirker_2019,WehnerStack1,WehnerStack2,WehnerStack3,ILLIANO2022109092,cacciapuoti2023quantum,Chung2022,Pirker2025}.

Building, engineering and operating a quantum network is a non-trivial task, as no single reference model yet exists \cite{Meter2013a,Pirker_2019,ILLIANO2022109092,Pirker2025,WehnerStack1,WehnerStack2,WehnerStack3,Pirker2025,Schon2024}. The aim of such a model is to organize quantum network devices in a model suitable for interoperable protocols. All proposals for reference models have in common that they will require some form of addressing. In classical networks devices use IP addresses to logically address other devices and to determine a path from the source to a destination. For quantum networks, not many works were done on how to facilitate addressing in quantum networks. The work of \cite{Miguel-Ramiro2021} considers using an address register to manipulate network states together with a shared, entangled task state. In \cite{cacciapuoti2023quantum} the situation with regards to quantum network addresses was summarized, and several open problems and questions were formulated. The recent work of \cite{Caleffi2025} was putting forward a scheme for quantum addressing utilizing quantum states as addresses to perform routing in quantum network devices. 

In this work we systematically introduce an addressing scheme that uses quantum states for entanglement-based quantum network. In such networks devices store pre-shared entanglement to respond quickly to incoming request. The network devices use the address state and the request state to apply, if selected, controlled quantum operations to a pre-shared network state. We note that only measurement outcomes must be classically communicated. Utilizing entanglement when distributing addresses opens a new research area for quantum networks. We consequently introduce a routing scheme that utilizes entanglement (as routing states) to perform controlled quantum teleporation \cite{Miguel-Ramiro2021} along a path in a network. Lastly, we prove that addresses using quantum states are equivalent to superposed tasks in a quantum network, and hence they offer a different view on the same concept. We note that the work here is one step towards a quantum control plane for quantum networks.

We start by relating to prior work in \ref{sec:background}. Section \ref{sec:addresses} presents our scheme on how to encode and address quantum network devices in a network. Section \ref{sec:routing} presents protocols for routing in a fully quantum manner. We discuss the equivalence of superposed tasks and quantum addresses in section \ref{sec:equi}. Finally we provide some applications of our scheme together with conclusions in section \ref{sec:applications} and section \ref{sec:conclusions}.

\section{Prior work}\label{sec:background}

In this work we use the Bell-state $\ket{\phi^{+}}$ which is defined as $\ket{\phi^{+}} = (\ket{00} + \ket{11})/\sqrt{2}$. When it comes to quantum addressing, there are only very few works on the subject \cite{VanMeter2022_InternetArch,Miguel-Ramiro2021,Caleffi2025}. In \cite{VanMeter2022_InternetArch} an architecture for the quantum internet was proposed using recursions based on classical addresses, similar in spirit to IP addresses. The work of \cite{Miguel-Ramiro2021} studies genuine quantum networks in which tasks are encoded by entangled quantum states shared between the network devices. The quantum network devices use such entangled states, also referred to as request states, to coherently execute operations on a shared, entangled, network state. This work also proposes and outlines first steps towards quantum addresses, however, it fails to fully discuss their utility. The work of \cite{Caleffi2025} systematically introduces quantum packets with headers encoding addresses, and presents a routing algorithm based on a variant of Grover's algorithm.

We extend the observations regarding quantum network addresses from \cite{Miguel-Ramiro2021,Caleffi2025}. First, in our scheme a request state passes through all network devices in a quantum network. This removes the necessity of having an entangled quantum state prepared beforehand to encode tasks \cite{Miguel-Ramiro2021}. Second, our protocol for executing operations as part of a request uses a full quantum description. Third, we propose a routing scheme for controlled teleportation using entanglement. Lastly, we  prove that quantum addressing in the sense we consider here is equivalent to executing tasks in superposition.

In this work we use the terms qubits and qudits interchangeably for convenience. When referring to a state $\ket{x}$ one can think of a collection of qubits implementing the state $\ket{x}$ in a system of dimension $2^d$ for some $d$.

\begin{figure}[h!]
\centering
\scalebox{2.0}{
\includegraphics{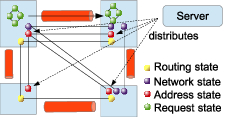}
}
\caption[h!]{\label{fig:overall} Network devices share three entangled states: An address state, a network state and a routing state. The request (and who participates) is encoded in another quantum state and passed from device to device. The devices use their address state and request state to decide if, and how, to change the network state. For controlled teleportation, they instead use address state, request state and routing state to perform controlled teleportation tasks.}
\end{figure}

The setting for this work is depicted and explained via Fig. \ref{fig:overall}

\section{Quantum addressing using quantum states}\label{sec:addresses}

\subsection{Quantum addresses}

We propose to encode the addresses of quantum network devices by a quantum state. We note that this may tightly couples different quantum network devices to each other in terms of entanglement. Formally, $n$ network devices share a state of the form $\sum_{i_1, \ldots, i_n} \beta_{i_1, \ldots, i_n} \ket{a_{i_1}}_1 \ldots \ket{a{_{i_n}}}_n$ with arbitrary weights $\beta_{i_1, \ldots, i_n}$ and $a_j$ denoting an address qubit (logical address) for $1 \leq j \leq n$. Such a state includes states in a tensor product (each device having an independent address) but also entangled state (correlated addresses), and we assume that a server distributes this state to the devices. To demonstrate the utility of entanglement in addresses, consider the following example. Suppose a quantum network comprises three devices sharing the address state $\frac{1}{\sqrt{3}}\left(\ket{1}_1\ket{2}_2\ket{3}_3 + \ket{2}_1\ket{3}_2\ket{1}_3 + \ket{3}_1\ket{1}_2\ket{2}_3\right)$. It is clear that this state decouples a logical address from physical devices of a network device due to entanglement. 

\subsection{Selecting a quantum network device}

For simplicity we assume that quantum network device $i$ receives the quantum state $\ket{a_i}$ where $a_i \neq 0$ as address state from a server. We denote the request state by $\ket{\psi_R}$, as described in Fig. \ref{fig:overall}, and it reads as $\ket{\psi_{R}} = \sum^{m}_{i=1} \beta_{i} \ket{b_i}$. It encodes the addresses of devices participating in the request via $\ket{b_i}$. The weights $\beta_i$ are arbitrary and depend on the request. To start request processing in the network the initiator of the request sends the request state $\ket{\psi_R}$ to the first quantum network device. When processing the request, this device performs a CNOT between its own address register and the request address register (referred to as selecting a device later). This results in the state $\sum^{m}_{i=1} \beta_{i} \ket{b_i} \ket{b_i \oplus a_1}$ where $\oplus$ denotes bit-wise XOR.

Device $1$ uses $\ket{b_i \oplus a_1}$ to apply controlled unitaries (classically communicated) to the pre-shared network state. After task processing completes, the device applies another CNOT to undo the effect of the first CNOT. After that, the device sends the resulting request state to the next device. After all parties have processed the request, the state of the network is $\ket{\psi_{R'}} = \sum^{m}_{i=1} \beta_{i} \ket{b_i} \ket{\psi_i}$ where $\ket{\psi_i}$ results from the controlled unitary of device $i$. We point out that when using entangled addresses in the network, this state will contain not only the superposition of individual operations and rather superpose also correlated operations in the constituents of an entangled state. 

\subsubsection{Encoding operations by classical means}

To encode the operation that a quantum network device should perform, the request initiator sends the classical descriptions of them to every device. Each quantum network device uses the state after selecting the device to apply the controlled unitary $((\id - \proj{0}) \otimes \id + \proj{0} \otimes U_j)$ where $U_j$ characterizes the required operation of the request.

\subsubsection{Encoding operations by quantum means}

Alternatively, one can encode the operations itself also in a quantum state by changing the request state $\ket{\psi_{R}}$ to 

\begin{equation} \label{eq:quantumopspre}
    \ket{\psi_{R}} = \sum^{m}_{i=1} \beta_{i} \ket{b_i} \ket{\varphi_{i,1}, \ldots, \varphi_{i,K}}.
\end{equation}

The values of $\ket{\varphi_{i,l}}$ encode for $1 \leq l \leq K$ the operations that quantum network $i$ should apply, chosen from a finite set. After the selection of device $j$, the resulting state of Eq. (\ref{eq:quantumopspre}) reads as

\begin{equation} \label{eq:quantumops}
    \sum^{m}_{i=1} \beta_{i} \ket{b_i} \ket{b_i \oplus a_j}_{j}\ket{\varphi_{i,1}, \ldots, \varphi_{i,K}}.
\end{equation}

To fulfill a task, we modify the controlled unitaries to 

\begin{equation}
    C^{(k)}_{ij} = ((\id - \proj{0\varphi_{i,j}})^{(0k)} \otimes \id + \proj{0\varphi_{i,j}}^{(0k)} \otimes U_j)
\end{equation}

where $k$ corresponds to the position of $\varphi_{ij}$ in Eq. (\ref{eq:quantumops}). The control on $\proj{0\varphi_{i,j}}^{(0k)}$ selects the state $\ket{\varphi_{i,j}}$ from the request state on position $k$. Network device $i$ must apply all possible controlled unitaries $C^{(k)}_{ij}$ for a $k$ to order apply the correct one, i.e. apply $C^{(k)}_i = \prod^m_{j=1} C^{(k)}_{ij}$. To process the entire request, device $i$ applies the unitary

\begin{equation}
    C_i = \prod^K_{k=1} C^{(k)}_i.
\end{equation}

To handle situations where not all devices need to apply the same number of operations we set the state $\ket{\phi_{\perp}}$ to encode the identity operation.

\section{Routing protocols for controlled teleportation}\label{sec:routing}

\subsection{A quantum routing protocol using local states}\label{sec:localrouting}

For this section we assume a network of Bell-states connecting quantum network devices in an arbitrary manner. The goal of routing is to select a path from a source device to a target device to teleport a qubit in a controlled manner \cite{Miguel-Ramiro2021}.

To perform local routing, each device of a network generates an entangled state locally that represents the network topology of Bell-pairs from its perspective, later used to perform controlled teleportation consuming neighboring Bell-pairs. For that purpose, device $j$ constructs a minimal spanning tree (MST) of the entire network, and for each path (in terms of Bell-states) in the network, the device applies the following rules:

\begin{itemize}
    \item Each node in the network gets represented by an address state $\ket{i}$ for device $i$.
    \item Each path without a junction is represented by a tensor product of the respective address states, i.e. $\ket{i}\ket{i+1}$ if device $i+1$ is located after device $i$ in the spanning tree of device $j$.
    \item Each time a path contains a junction device $j$ represents the junction by summing over the address state, i.e. $(\ket{i} + \ldots + \ket{i+m})$ for $m$ branches.
\end{itemize}

It must be noted that all qubits here reside at device $j$, and hence no sub-index is necessary. Furthermore, we omit normalization and weights in the superposition in this entire section for clarity. We also note that an MST does not contain cycles because its a tree by construction.



\begin{figure}[h!]
\centering
\scalebox{1.7}{
\includegraphics{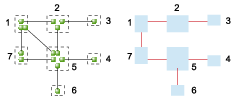}
}
\caption[h!]{\label{fig:routing_local_tree} A quantum network and its associated spanning tree for routing from the perspective of device $1$.}
\end{figure}

For illustration, consider the network shown in Fig. \ref{fig:routing_local_tree}. It shows a network of $7$ devices. The routing state of device $1$ reads as $\ket{2}\ket{3}\ket{\perp} + \ket{7}\ket{5}(\ket{4} + \ket{6})$. We note that the state $\ket{\perp}$ is necessary to ensure the dimensions of the states are correct. This state resembles a spanning tree of the network. It encodes which Bell-state must be used by device $1$ for controlled teleportation. The state reflects the path $2 \to 3$ as well as $7 \to 5 \to 4$ and $7 \to 5 \to 6$, and these constituents encode which Bell-state device $1$ must use for teleportation for a given target node. 

Formally, to route a qubit using teleportation and a routing state, device $j$ uses controlled unitaries (together with Bell-state measurements) of the form

\begin{align}
    C_{k,t,l} = ((\id - \proj{k} & \otimes \proj{t}^{(l)}) \otimes \id \notag \\
    & + \proj{k} \otimes \proj{t}^{(l)} \otimes S_{kj})
\end{align}

where $S_{kj}$ denotes the operation to implement a controlled teleportation using $\ket{\phi^{+}}_{kj}$. The protocol for controlled teleporation follows \cite{Miguel-Ramiro2021}, including correction operations. Network device $j$ applies the unitaries $C_{k,t,l}$ for all $k$ that are connected neighbors to device $j$ to select a particular Bell-pair coherently for controlled teleportation. The parameter $t$ corresponds to the address of the target device of the request, and is fixed. Device $j$ then runs $C_{k,t,l}$ for all $k$ and all $l$, where $l$ takes all values till the number of quantum systems for encoding the routing state are used up. In simple words, looping through all $l$ for a fixed $k$ ensures that if a path contains $t$, it will get chosen for teleportation.


One can simplify the routing state to $\left( \sum_j \ket{\mathrm{reach}(\ket{\phip}_{ij})} \right)$ where $\ket{\mathrm{reach}(\ket{\phip}_{ij})}$ denotes a tensor product of all quantum address states reachable via Bell-state $\ket{\phip}_{ij}$ in terms of controlled teleportation, e.g. $\ket{2}\ket{3}\ket{\perp}\ket{\perp} + \ket{4}\ket{5}\ket{6}\ket{7}$, since devices $2,3$ are reachable by $\ket{\phi^+}_{12}$ and devices $4,5,6,7$ through $\ket{\phi^+}_{17}$.

\subsection{A quantum routing protocol using entangled states}


For distributed routing, each network device creates a MST from its perspective of the Bell-state topology, as also in the first part of the local routing protocol in section \ref{sec:localrouting}, see also Fig. \ref{fig:routing_local_tree}.

\begin{figure}[h!]
\centering
\scalebox{1.85}{
\includegraphics{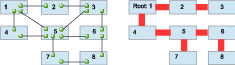}
}
\caption[h!]{\label{fig:spanning_tree} A network of eight devices, where channels represent Bell-states for teleportation, together with a minimal spanning tree (MST) for device $1$.}
\end{figure}

Each network device (source node) creates an entangled state reflecting the topology of the MST of its perspective. These entangled states (MST states) encode for each node the routes to reach other devices from the perspective of the source node, and they distribute the qubits accordingly in the network. The construction scheme for the states follow the rules outlined in section \ref{sec:localrouting}. For example, device $1$ can reach through network device $5$ in Fig. \ref{fig:spanning_tree} either device $7$ or devices $6$ and $8$. Therefore, the portion of the MST state for source node $1$ of device $5$ corresponds to $\ket{7,\perp}_5 + \ket{6,8}_5$, where $\ket{\perp}$ encodes a dummy state. Applying this idea iteratively, starting from source device $1$ results in the MST state

\begin{align}
    \ket{2,3,\perp}_1 & \ket{3}_2 + \notag \\
    & \ket{4,5,6,7,8}_1\ket{5,6,7,8}_4(\ket{7,\perp}_5 + \ket{6,8}_5\ket{8}_6)
\end{align}

We observe that this state encodes which device is reachable through a particular Bell-state of the network. For example, the state $\ket{2,3,\perp}_1$ of device $1$ encodes the Bell-state connecting device $1$ and $2$, because network device $1$ can reach devices $2$ and $3$ by it. Similarly, the state $\ket{5,6,7,8}_4$ encodes the Bell-state between device $4$ and device $5$ since device $4$ can reach the devices $5, 6, 7$ and $8$ by consuming it. Generally speaking, the state $\ket{\mathrm{reach}(s,\ket{\phip}_{ij})}_i$ denotes all the destination addresses that can be reached by consuming the Bell-state $\ket{\phip}_{ij}$ from the perspective of a source node $s$. The algorithm uses the recursion construction

\begin{align}
    \ket{\mathrm{reach}(s,i)}_i = \sum_j \ket{\mathrm{reach}(s,\ket{\phip}_{ij})}_i \otimes \ket{\mathrm{reach}(s,j)}_j
\end{align}

The function $\mathrm{reach}(s,j)$ recursively establishes the state for routing. We note that in contrast to section \ref{sec:localrouting} the state here is distributed. To be able to perform transmissions among all devices, each network device creates such a MST state. 

We achieve routing by using the MST state for controlled teleportation from a source to a destination by encoding the destination address into an address state and using the routing state to select a path. We observe that the network devices not only teleport the state to be send but also the target address qubit to the next hops of the teleportation.

The MST states of each device can also be unified into a single state by either encoding more routing information at each device, or by superposing the individual MST states in a superposition encoding the source devices $\ket{R} = \sum_i \ket{R_i} \otimes \ket{i}^{\otimes n}$ where $\ket{R_i}$ corresponds to the MST state of device $i$, and each of the devices receives one state of $\ket{i}^{\otimes n}$.

The number of systems required for a single MST state is at most $n!$ (1D cluster state), and the number of systems for overlaying them is  $n \cdot n!$ (because of $\ket{i}^{\otimes n}$ in $\ket{R}$). The main complexity of this schemes originates from the preparation of the MST states, as routing itself is rather straightforward to implement via controlled unitaries. To simplify routing, and also reduce storage, one can combine local routing scheme and distributed routing into a hybrid approach to reduce the complexity of preparing and distributing MST states.

\section{Equivalence of superposed addresses and superposed tasks}\label{sec:equi}

In this section we prove that superposed addresses are equivalent to superposed tasks from a theoretical perspective. For that purpose, consider a request state of the form 

\begin{align}
    \ket{\psi_R} = \sum^n_{i=1} \alpha_i \ket{i} \ket{\varphi_{i,1}, \ldots, \varphi_{i,m}}.
\end{align}

The values of $\ket{i}$ correspond to the addresses of the devices that need to execute $\ket{\varphi_{i,1}, \ldots, \varphi_{i,m}}$. Further assume that the network is in the following address state

\begin{align}
    \ket{\psi_A} = \sum^n_{j=1} \beta_j \ket{a_{1,j}, \ldots, a_{n,j}}_{1,\ldots,n}.
\end{align}

This corresponds to the most general address state of a network, and includes address states in tensor product but also entangled addresses. Further we assume that the network connects through a network state $\ket{\psi_N}$. The overall state reads as

\begin{align}
    & \ket{\psi_R} \ket{\psi_A} \ket{\psi_N} = \notag \\
    & = \sum^n_{i,j=1} \alpha_i \beta_j \ket{i} \ket{a_{1,j}, \ldots, a_{n,j}}_{1,\ldots,n} \ket{\varphi_{i,1}, \ldots, \varphi_{i,m}} \ket{\psi_N}. \label{eq:equi:1}
\end{align}

The network devices use the result of the CNOT between the states $\ket{i}$ and $\ket{a_{k,j}}$ (selecting a device) to determine if the operations encoded by $\ket{\varphi_{i,1}, \ldots, \varphi_{i,m}}$ require execution or not. By setting $\ket{t_{i,j}} = \ket{i} \ket{a_{1,j}, \ldots, a_{n,j}}_{1,\ldots,n}$ and $\gamma_{i,j} = \alpha_i \beta_j$ Eq. (\ref{eq:equi:1}) reads as

\begin{align}
    \ket{\psi_R} \ket{\psi_A} \ket{\psi_N} = \sum^n_{i,j=1} \gamma_{i,j} \ket{t_{i,j}} \ket{\varphi_{i,1}, \ldots, \varphi_{i,m}} \ket{\psi_N}.
\end{align}

This state is very closely related to the superposed task state used in \cite{Miguel-Ramiro2021}. By replacing the indices $i,j$ with $k$, and equipping $\ket{t_{i,j}}$ as well as the program state with a tensor product $n$, we end up with a superposed task state according to \cite{Miguel-Ramiro2021}, i.e.

\begin{align}
    \ket{\psi_R} \ket{\psi_A} = \sum^n_{k=1} \gamma_{k} \ket{t_{k}}^{\otimes n} \ket{\varphi_{k,U}}^{\otimes n}.
\end{align}

We observe that in this situation we just need to pass to each device the information about the branches $\ket{t_{k}}$ where the device shall participate. Hence, from a theoretical perspective, quantum addresses and superposed tasks are equivalent. 

\section{Applications}\label{sec:applications}

We move on to some simple applications of the proposed addressing besides the controlled teleportation outlined in the section \ref{sec:routing}.  Suppose a network of three devices require sharing different network states depending on an address configuration. We label these network states by $\ket{N_1}$, $\ket{N_2}$ and $\ket{N_3}$. In this case, an entangled address state allows to attach a network state to an address configuration. That is, the devices could share an overall state of the form

\begin{align}
    \alpha_1 \ket{1}_1 & \ket{2}_2 \ket{3}_3\ket{N_1}_{123} \notag \\
    & + \alpha_2 \ket{2}_1\ket{3}_2\ket{1}_3\ket{N_2}_{123} + \alpha_3 \ket{3}_1\ket{1}_2\ket{2}_3 \ket{N_3}_{123} \label{eq:noverlay}
\end{align}

For such a state, the network state to which operations apply depends on the address configuration assigned to the devices. Therefore, the addressing scheme proposed here enables quantum network devices to use different resources for accomplishing tasks in a network depending on the address configuration. Furthermore we observe that the quantum network state that associates with a given address configuration may have completely different entanglement topologies. It was shown in \cite{Miguel-Ramiro2021} that such schemes may even preserve entanglement in case of losses. We highlight another interesting application: Protocols can be executed on different network states (one per address configuration), but yet leading to the same output state. For example, in the situation of Eq. (\ref{eq:noverlay}) the protocols executed in the individual branches on $\ket{N_1}$, $\ket{N_2}$ and $\ket{N_3}$ could result in a common state $\ket{G}$, which keeps the address state intact and reusable.

\section{Conclusions}\label{sec:conclusions}

In this work we have presented an approach to address devices in a quantum network using quantum states. We proposed the use of such quantum states to encode the address of a device where each device receives its address from a quantum address server. The addresses of quantum network devices may either be in a product state, or an arbitrary entangled state. The latter scheme appears very interesting from several perspectives, because it allows to overly different address configurations together with different network states in a quantum network. We discussed how to address and select devices based on their addresses. From there we continued to put forward two proposals to perform routing for controlled teleportation in a quantum network, both in a local way but also a distributed way by using MST states. As additional applications we highlighted potential applications to the entanglement-based quantum networks.


\section{Acknowledgments}  
The authors thank J. Chung for constructive feedback on this manuscript. This research was funded in whole or in part by the Austrian Science Fund (FWF) 10.55776/P36010. For open access purposes, the author has applied a CC BY public copyright license to any author accepted manuscript version arising from this submission.




\bibliographystyle{IEEEtran}
\bibliography{quantum_networks}

\end{document}